\documentstyle[aaspptwo]{article}     
\def\cm{~cm$^{-1} $}
\def\mic{$\mu$m}
\def\etal{et al.\/ }
\def\4f{$3d^64s(^6D)4f$}
\def\5g{$3d^64s(^6D)5g$}
\def\D4d{$3d^64s(^6D)4d$}
\begin{document}
\title{Analysis of the $3d^64s(^6D)\ 4f-5g$ supermultiplet of Fe~I
            in laboratory and solar infrared spectra  }
\author{
S.~Johansson\altaffilmark{1} \altaffilmark{2}
G.~Nave     \altaffilmark{1} \altaffilmark{3}
M.~Geller   \altaffilmark{4}
A.~J.~Sauval \altaffilmark{5}
N.~Grevesse \altaffilmark{6}
W.~G.~Schoenfeld \altaffilmark{7} E.~S.~Chang \altaffilmark{7}
and
C.~B.~Farmer \altaffilmark{4}
}
\altaffiltext{1}{Department of Physics, University of Lund, S\"olvegatan 14,
S-223 62 Lund, Sweden}
\altaffiltext{2}{Lund Observatory, Box 43, S-221 00 Lund, Sweden}
\altaffiltext{3}{Blackett Laboratory, Imperial College,
                 Prince Consort Rd, London, U.K. SW7 2BZ}
\altaffiltext{4}{ Jet Propulsion Laboratory, 4080 Oak Grove
Drive, Pasadena, CA 91109, U.S.A.}
\altaffiltext{5}{ Observatoire Royal de Belgique, Avenue
                  Circulaire 3, B-1180 Brussels, Belgium}
\altaffiltext{6}{ Institut d'Astrophysique, Universit\'e de
                   Li\`ege, Avenue de Cointe 5, B-4000 Li\`ege, Belgium}
\altaffiltext{7}{ Dept. of Physics and Astronomy,
        University of Massachusetts, Amherst, MA 01003, U.S.A }

\begin{abstract}
The combined laboratory and solar analysis of the highly-excited
subconfigurations \4f and \5g of Fe~I has allowed us to classify 87
lines of the $4f-5g$ supermultiplet in the spectral region 2545-2585~\cm.
The level structure of these JK-coupled configurations is predicted
by semiempirical calculations and the quadrupolic approximation.
Semiempirical gf-values have been calculated and are compared to gf
values derived from the solar spectrum. The solar analysis has shown
that these lines, which should be much less sensitive than lower excitation
lines to departures from LTE and to temperature uncertainties, lead to a
solar abundance of iron which is consistent with the meteoritic value
($A_{Fe} = 7.51$).
\end{abstract}
\vspace{1cm}
\begin{center}
{\large To appear in the Astrophysical Journal}
\end{center}

\section{Introduction}

The spectrum of Fe~I is a significant opacity source in spectra of the
sun and cool stars, and recent studies have illustrated the importance
of highly excited configurations in the infrared spectrum of the sun. On
the basis of a laboratory Fourier Transform (FT) spectrum Bi\'emont
\etal (1985) published a list of $\sim$1200 iron lines between 1~\mic\
and 4.1~\mic\ that corresponded to lines in the solar spectrum. Roughly
300 of these lines had been identified in FT spectra by Litz\'en and
Verg\`es (1976), and another 370 were observed in a very recent study of
highly excited configurations in Fe~I (Nave \& Johansson 1993). The \4f
subconfiguration was established from $4d-4f$ transitions by Johansson
\& Learner (1990), which resulted in the identification of about 200
solar lines in the atmospheric windows between 1.4~\mic\ and 2.1~\mic.

In this paper we present the analysis of the \5g subconfiguration of
Fe~I that has been established by transitions to the \4f levels
(Johansson \& Learner 1990) using laboratory FT spectra. Since the
excitation energies of the \5g and \4f subconfigurations are only 0.5 to
0.8~$eV$ lower than the ionization energy, 7.9~$eV$, these very high
excitation lines of Fe~I provide important information about plasma
conditions. Even if there are slight non-LTE effects in Fe~I for levels
of lower excitation energies (Blackwell \etal 1984), lines from high
excitation levels should be formed in LTE and should be rather
insensitive to slight temperature uncertainties. The $4f$ and $5g$
electrons do not significantly penetrate the $3d^64s$ core, and the
positions of the \4f and \5g subconfigurations relative to the series
limit can therefore be predicted. Both subconfigurations are best
described in the JK coupling scheme. The $4f-5g$ lines occur in a narrow
region of 35~\cm\ (0.056~\mic), 100~\cm\ to the blue of the Br $\alpha
(4-5)$ line of hydrogen at 2467.75~\cm\ (4.05~\mic). This region is
shown in Fig. 1.

The presence of $4f-5g$ lines in ground-based IR spectra of the sun and
of $\alpha$ Tau (Ridgway \etal 1984) was first reported by Johansson
\etal (1991). Most of the high resolution infrared atlases of the solar
spectrum have been obtained from ground-based observations, and in some
regions the telluric contribution totally obscures the lines.
The solar spectrum has been recently obtained with the ATMOS space
experiment (Farmer \& Norton 1989). It is completely free of telluric
lines, and the high resolution and signal-to- noise ratio allow us to
make a much more detailed analysis of the Fe~I $4f-5g$ lines.
Preliminary results have already been given by Geller (1992), Johansson
\etal (1993), and Schoenfeld \etal (1993a).

\section{Laboratory Observations and Analysis}

The laboratory spectrum used was recorded on the Fourier transform spectrometer
at the National Solar Observatory, Tucson, and details of the experimental
setup have been described elsewhere (Learner \& Thorne 1988). The source
was a hollow cathode of pure iron run in 3.7 Torr of neon and a DC
current of 1.4~A. The wavenumber resolution is 0.012~\cm. The $4f-5g$
transitions reported here are also present at a lower signal-to-noise on
other FT spectrograms taken in this region. The better quality of the
spectrogram used in the present study appears to be mainly due to water
cooling of the cathode in the light source.

The spectra were reduced with the DECOMP program of Brault and Abrams
(1989) to give the wavenumber, integrated intensity, width and damping
parameter for each line. The spectra were wavenumber calibrated from 26 Ar
II lines (Norl\'en 1973) present in our visible region spectra, and the
calibration was
carried into the infrared by using overlapping wide-range spectra (Nave
\etal 1992). The calibration constant for the spectrum used in this
study was less than 0.001~\cm. The strongest, unblended lines in the
spectrum are determined with a precision of $\sim$0.0005~\cm, but almost
all of the $4f - 5g$ lines are blended. The uncertainty of weaker lines
is given by the full width at half maximum divided by the signal-to-noise
and can be as much as 0.008~\cm\ for the weakest lines.

Both the \4f and \5g subconfigurations are best represented in the JK
coupling scheme (Johansson \& Learner 1990). The gross structure
within each subconfiguration is defined by the fine structure splitting of
the $3d^64s\ ^6D$ term in
Fe II. The energy levels are then split into pairs, and the distribution
of the centres of gravity of the pairs is described by a quadrupolic
approximation, determined by the electrostatic parameter $F^2(3d,nl)$.
If the centres of gravity are plotted against the scalar product $h=J_c\cdot l$
of
the total angular momentum of the core $J_c$ and the orbital angular
momentum of the outer electron $l$, a set of parabolas is obtained --
one for each value of $J_c$. As previously noted for
the $^5D$ term of $3d^6$ (Johansson \& Learner 1990), both bowl-shaped
and umbrella-shaped parabolas are obtained. This is due to a change of the
sign in the energy expressions for JK-coupled subconfigurations built on
particular Hund terms, which will be thoroughly discussed in a forthcoming
paper (Schoenfeld \etal 1993b). The vertices of the parabolas are predicted
to be at $h = -\frac{1}{4}$. In Fig. 2 we have fitted parabolas to the
observed energy levels of \5g. The experimentally determined vertices
lie from $h = -0.01$ to $0.05$, and the deviations of the energy levels
from the parabolic curve are $\sim 0.02$\cm. In  the \4f subconfiguration the
energy levels are, with one exception (see below), split into resolved pairs
by the Coulomb exchange interaction, but in \5g the pair splittings are
too small to be resolved. Values for each energy level of a pair have been
determined from lines in which theoretical gf-values predict only one
transition to be important.

The experimentally determined energy levels are given in Table 1. The
majority have been determined from the laboratory line list presented in
Table 2. However, a few energy levels with low J have not been found as
the necessary lines were not present in our laboratory spectra.
Calculated gf-values of lines from these levels imply that they would
not be distinguishable from the noise. The positions of these ``missing''
energy levels have been predicted from the parabolas in Fig. 2 and the
predictions agree with a quadrupole moment polarization theory
(Schoenfeld \etal 1993a). Many lines from these predicted energy levels
appear weakly in the solar spectrum, and this has enabled us to
establish another three levels which are marked with the superscript
`S' in Table 1. The remaining two predicted levels are marked with the
superscript `P' and are given to two decimal places. Two decimal places
are also given if both levels of a pair are determined by a
line in the laboratory spectrum from which only
one wavenumber is determinable. An example is the line at 2556.948~\cm\
which is the only line determining the pair of levels
$5g(\frac{9}{2})[\frac{17}{2}]$. In this case neither the  $4f$ nor
$5g$ pair splitting can be resolved. If a level is determined from weak
or blended lines, for which the wavenumbers are uncertain, it is also
given to two decimal places (e.\/g.\/ $5g(\frac{1}{2})[\frac{9}{2}]$).

We have revised the energy levels of both \4f and \5g by using a least
squares fitting program, and have connected them to the rest of the term
diagram of Fe I by means of transitions to \D4d. The energy level values
we have obtained for the \4f levels are systematically higher than those
reported previously (Johansson \& Learner 1991) by $\sim 0.004$~\cm,
since the wavenumbers for $4d - 4f$ transitions in our other set of
infrared laboratory spectra are also systematically higher than
previously reported. We attribute this difference to the improved method
of wavenumber calibration that we have used in the current spectra.
Estimates of the accuracy of the levels are given in column 4 of Table
1, and range from 0.003~\cm\ for levels determined from strong, unblended
lines to 0.01~\cm\ for levels determined from blended lines. Predicted
levels are probably accurate to 0.05~\cm. However, it should be noted
that these energy levels refer to one set of plasma and discharge
conditions in a hollow cathode lamp, and deviations of several mK (1mK =
0.001\cm) may be found when comparing them with data obtained under
other operating conditions or from other light sources. For instance, a
systematic shift of $\sim$0.006~\cm\ is observed between the laboratory
and solar data used in this study, where the laboratory wavelengths are
about 0.1~\AA\ longer than the solar wavelengths. This is further
discussed in Sec.3.

The similar arrangement of the energy levels in \4f and \5g means that
all the $4f-5g$ lines are concentrated into a narrow region from
2545~\cm\ to 2580~\cm. Almost all of the laboratory lines in this region
are due to this supermultiplet, and unblended lines have a consistent
width of 0.045 -- 0.050~\cm.  All of the laboratory lines are also
present in absorption in the ATMOS spectrum of the sun, as can be seen
from Fig. 1. The identified $4f-5g$ lines are presented in Table 2.
Table 3 gives lines observed in the solar spectrum but not in our
laboratory spectra that have been identified as $4f-5g$ lines by using
the energy levels of table 1. As previously mentioned, some of these are
too weak to be distinguishable from the noise in the laboratory
spectrum, but others are strong and are masked by neon lines.

The reliability of the identifications in Table 2 can be seen by comparing the
intensities of lines measured in the laboratory and solar spectra (Fig. 3).
The large scatter at low intensities suggests there are still unresolved
blends or incorrectly fitted lines in both the laboratory and solar
spectrum; we note also that the fit of the weakest solar lines is generally
less accurate than that for other lines.
The points with the largest deviation correspond to
regions where the blending is particularly severe ($\sim$2563.6~\cm\ and
2567.8~\cm). Points lying below the line indicate a blend with another
iron line in the laboratory spectrum. Points lying above the line
indicate blends in the solar spectrum and an example of such a blend
with a Si line is marked. Moreover, the curve of growth effect, which is
well visible in our strongest solar lines, clearly explains the
non-linear relation between laboratory and solar intensities (see also
section 3).

The ninth column of Table 2 gives log(gf) values for the \4f $-$ \5g
supermultiplet which have been calculated with the Cowan computer code
(Cowan 1981). Most of them agree to within 1\% with the hydrogenic gf-values
calculated in the pure JK coupling scheme. The calculations
predict that the strongest transitions are of the form $(J_c)[K]_J -
(J_c)[K+1]_{J+1}$, and that no transitions with $\Delta J_c \neq 0$ are
expected. The few differences between the calculations and the
hydrogenic gf-values can be traced to mixing between different $4f$
levels. For example, calculations predict that the
$4f(\frac{7}{2})[\frac{3}{2}]_2$ at 57152.412~\cm\ and
$4f(\frac{7}{2})[\frac{5}{2}]_2$ at 57154.265~\cm\ are mixed, and an
examination of the $4f - 5g$ intensities suggests the designations of
these two levels should be exchanged. This is confirmed by a better fit
of the revised levels to the parabolic curve for the $J_c=(\frac{7}{2})$
levels of $4f$. However, even if the designations are exchanged the
calculated gf-values for all transitions involving these two levels are
less reliable. This level mixing enables $\mid \Delta K \mid = 2$
transitions to occur. For example, in Table 2 the line at 2567.683~\cm\
arises from K=3.5 $\rightarrow$ 5.5, and in Table 3 the line at
2563.052~\cm\ arises from the K=1.5 $\rightarrow$ 3.5 transition.

An example of mixing between a $4f$ level and another Fe~I level is
illustrated by  $4f(\frac{5}{2})[\frac{5}{2}]_2$ at 57431.116~\cm, which
is mixed with $3d^64s(^6D)6p\,^5\!P_2$ at 57437.192 \cm. This results in
a line at 2562.206~\cm\ due to the transition
$6p\,^5P_2 - 5g(\frac{5}{2})[\frac{5}{2}]_2$. The transition $6p\,^5P_2
- 5g(\frac{5}{2})[\frac{7}{2}]_3$ contributes to the blend in the solar
spectrum at 2563.52~\cm\ and is probably the reason for the fact that
$4f(\frac{5}{2})[\frac{5}{2}]_2  - 5g(\frac{5}{2})[\frac{7}{2}]_3$ is
not observed. As this mixing is not predicted by calculations it is not
possible to calculate gf-values for these transitions, and the calculated
gf-values for transitions involving $4f(\frac{5}{2})[\frac{5}{2}]_2$ may
be too large.

\section{Solar Analysis}

Between 2545 and 2585~\cm\, more than ninety percent of the solar lines
in the ATMOS infrared solar spectrum (Farmer \& Norton 1989) are identified
as Fe~I $4f-5g$ transitions. These relatively faint lines in the solar
spectrum have profiles that are typical for highly-excited transitions:
they are broad with very extended wings. They have almost Lorentzian
shapes whereas profiles of faint, low-excitation lines are Gaussian.

Although faint, these 4f-5g lines are very sensitive to the damping constants.
This was shown in Fig. 3 of Johansson \etal (1993), where we
only considered broadening by collisions with hydrogen atoms, computed
from the Uns\"old (1955) approximation. This frequently used approximation
should, in principle, be valid for these highly excited ``hydrogenic'' levels.
We found, however, that the damping constant computed in this way was too
small to reproduce the observed solar lines profiles. This is also the case
for lines between levels of low excitation (see e.g. Blackwell \etal 1984;
Holweger \etal 1991). As shown in Fig. 4, very good agreement
is obtained between synthetic and observed line profiles when the collisional
damping constant is multiplied by an enhancement factor of about 2.5.

Even if the broadening due to collisions with neutral particles increases
tremendously with the excitation energy, the Stark broadening increases
even more rapidly. For the relatively high excitation energy associated with
the Fe~I $4f-5g$ lines, the Stark broadening ( Chang \& Schoenfeld 1991;
Carlsson \etal 1992) is only about 1/3 of the Van der Waals
broadening. For transition arrays at even higher excitation energies, e.g.
the $5g-6h$ lines of Fe~I (Schoenfeld \etal 1993b), the Stark broadening
becomes
more important.

We have fitted synthetic line profiles to the observed solar spectrum in
the region of the Fe~I $4f-5g$ lines between 2545 and 2585~\cm\ using the
photospheric model of Holweger \& M\"uller (1974), together with a
microturbulent velocity of 1 km/s. An example of our fit is shown in
Fig. 5, together with the relevant portion of the laboratory spectrum.
We have used a solar iron abundance of $A_{Fe} = 7.51$
(in the usual scale, $A_{Fe} = \mbox{log}(N_{Fe}/N_H)+12.0$), which is
the meteoritic value (Anders \& Grevesse 1989). Recent investigations by
different authors (Holweger \etal 1990; Holweger \etal 1991; Bi\'emont
\etal 1991; Hannaford \etal 1992) have shown that the photospheric
abundance of iron agrees with the meteoritic value.

We have been forced to apply a somewhat larger Doppler correction to the
solar $4f-5g$ Fe~I lines than in the case of CO lines (see Fig. 5 in
Farrenq \etal 1991) and of low excitation Fe~I lines. The solar
wavenumbers of our $4f-5g$ Fe~I lines, which are given in the fifth
column of table 2, have been corrected by a total Doppler shift of 66~mK
instead of about 60~mK as found for other low excitation solar lines. We
are unable to attribute this additional shift of about 6~mK to conditions
in the solar photosphere. It is more likely to be due to pressure
shifts in the hollow cathode; calibration problems
would also affect the low excitation lines (Nave \etal 1992).
Stark shifts in the hollow cathode, which can be as large as 30~mK in
certain lines of neon (Chang \etal 1993), are calculated to be less than
1~mK for the $4f-5g$ lines of Fe I.

Our ``astrophysical gf-values'', derived from the ATMOS solar spectrum and the
meteoritic iron abundance, are given in the seventh column of Table 2.
The best solar line profiles are marked by an asterisk in the last column.
Some additional solar lines (such as Si~I, CH and OH and other, as yet
unidentified lines from Geller (1992)) were simulated through
artificial metallic lines and fitted to the observations.
``Astrophysical gf-values'' for lines not present in the laboratory
spectrum are given in the fifth column of Table 3 together with
calculated gf-values in column 7.

Good agreement is found between solar-derived gf-values and the calculated gf
values, as can be seen from Fig. 6 where $\log(gf)$ is plotted against
the log of the calculated gf-value. Most of the points lie on a
well-defined line, which proves the consistency of the set of calculated
gf-values. Based on the best selected solar lines, the solar-derived gf
values are systematically somewhat smaller than the calculated values
by a mean amount of about -0.10$\pm$0.05 dex. Although this could be due to the
adopted solar abundance of iron, the difference is more likely due to
uncertainties both in the calculated and solar gf-values.
The main uncertainties of our solar-derived gf-values are due to the adopted
damping constants (which require a rather large enhancement factor of
about 2.5) and to an uncertainty in our measured intensities (due to
a possible error of a few percent in the zero intensity level and also to the
difficulty of determining the true local continuum in the solar spectra).
The discrepancy between solar and theoretical gf values quoted
hereabove is the highest possible value obtained when adopting the
lowest possible continuum value. Therefore the general agreement remains
within all experimental and theoretical combined errors.

It is interesting to compare the \4f - \5g supermultiplet in Fe~I with the
$3s4f-3s5g$ multiplet in Mg~I, for which extensive solar atmosphere modelling
has
been carried out (Chang \etal 1992). There, the hundred or so lines in
only one Fe~I supermultiplet are reduced to a few blended lines in one
$^1F- ^1G$ and one $^3F- ^3G$ multiplet in Mg~I. The Mg
line cores were found to be formed at a height of 350 km (above
$\tau_{500}$=1). Since Mg and Fe have comparable ionization potential
and abundance, the weaker Fe I lines are expected to be formed lower in
the solar atmosphere. Our calculations show that the center of the
strongest  Fe $4f-5g$ lines are formed at a height of $\sim$240~km, with
the weaker lines formed deeper ($\sim $175~km) in the solar atmosphere,
at about the same height as the fainter Mg I lines (e.g. $5f-7g$).

\section{Conclusions}

We have classified 87 lines of the Fe~I \4f$-$\5g
supermultiplet by combining laboratory and solar data with theory. We
have established all but two of the 58 levels in the \5g
subconfiguration, 53 from laboratory FT spectra, and three from the
solar spectrum. Both configurations are best described using the JK
coupling scheme. All the classified lines strictly obey the
$\Delta J_{c}=0$ rule, but we have found some violations of the other
rule, $\mid \Delta K \mid \leq 1$. These are caused by mixing of certain
levels of the same $J_{c}$ and J, but of different K in the $4f$
subconfiguration. Besides this level mixing, the $4f$ subconfiguration
is also subject to configuration interaction with $3d^64s(^6D)6p$, due
to the accidental coincidence of $6p\ ^5\!P_2$ level and $4f\
(\frac{5}{2})[\frac{5}{2}]_2$. The 5g configuration is found to be in JK
coupling without any mixing whatsoever.

All of the lines in the \4f$-$\5g supermultiplet that we have observed
in the laboratory are also present in the solar spectrum and account for
more than 90~\% of the lines between 2545 and 2585~\cm. We have calculated
semi-empirical gf-values for most of the identified lines and have compared
these with gf-values derived from the solar spectrum. From these $4f-5g$
lines, which we expect to be much less sensitive than lower excitation
lines to departures from LTE and to temperature uncertainties, we have
determined a solar abundance of iron which is consistent with the
meteoritic value ($A_{Fe} = 7.51$).

\acknowledgments

We are grateful to R.~C.~M.~Learner and J.~W.~Brault for providing
the laboratory FT data for this work. We also want to thank W.~Nijs (Brussels)
for his help in all calculations of solar spectra. G.~Nave gratefully
acknowledges an external fellowship from the European Space Agency,
S.~Johansson financial support from the Swedish National Space board,
and N.~Grevesse financial support from the Belgian Fonds National de
la Recherche Scientifique.

\onecolumn
\clearpage
\begin{planotable}{llll}
\tablewidth{0pt}
\tablecaption{New $3d^64s(^6D)5g$ energy levels of Fe~I}
\tablehead{ \colhead{designation\tablenotemark{a}}  &
            \colhead{~J~}                           &
            \colhead{Level }                        &
            \colhead{Error }                        \\
            \colhead{(abbr)}                        &
            \colhead{ }                             &
            \colhead{($cm^{-1}$)}                   &
            \colhead{($cm^{-1}$)}                   }
\startdata
 $(4.5) [8.5] $& 9 &   59335.73    & .01        \nl
 $(4.5) [8.5] $& 8 &   59335.72    & .01    \nl \nl
 $(4.5) [7.5] $& 8 &   59331.269   & .005       \nl
 $(4.5) [7.5] $& 7 &   59331.270   & .004   \nl \nl
 $(4.5) [6.5] $& 7 &   59329.645   & .004       \nl
 $(4.5) [6.5] $& 6 &   59329.642   & .004   \nl \nl
 $(4.5) [5.5] $& 6 &   59329.892   & .004       \nl
 $(4.5) [5.5] $& 5 &   59329.891   & .003   \nl \nl
 $(4.5) [4.5] $& 5 &   59331.286   & .004       \nl
 $(4.5) [4.5] $& 4 &   59331.287   & .004   \nl \nl
 $(4.5) [3.5] $& 4 &   59333.255   & .004       \nl
 $(4.5) [3.5] $& 3 &   59333.257   & .004   \nl \nl
 $(4.5) [2.5] $& 3 &   59335.316   & .003       \nl
 $(4.5) [2.5] $& 2 &   59335.317   & .005   \nl \nl
 $(4.5) [1.5] $& 2 &   59337.081   & .005       \nl
 $(4.5) [1.5] $& 1 &   59337.078   & .005   \nl \nl
 $(4.5) [0.5] $& 1 &   59338.255   & .01        \nl
 $(4.5) [0.5] $& 0 &   59338.27    & .01    \nl \nl
 $(3.5) [7.5] $& 8 &   59717.941   & .007       \nl
 $(3.5) [7.5] $& 7 &   59717.94    & .01    \nl \nl
 $(3.5) [6.5] $& 7 &   59717.084   & .005       \nl
 $(3.5) [6.5] $& 6 &   59717.079   & .004   \nl \nl
 $(3.5) [5.5] $& 6 &   59716.793   & .004       \nl
 $(3.5) [5.5] $& 5 &   59716.792   & .003   \nl \nl
 $(3.5) [4.5] $& 5 &   59716.936   & .004       \nl
 $(3.5) [4.5] $& 4 &   59716.944   & .004   \nl \nl
 $(3.5) [3.5] $& 4 &   59717.325   & .004       \nl
 $(3.5) [3.5] $& 3 &   59717.317   & .004   \nl \nl
 $(3.5) [2.5] $& 3 &   59717.764   & .004       \nl
 $(3.5) [2.5] $& 2 &   59717.759   & .004   \nl \nl
 $(3.5) [1.5] $& 2 &   59718.13    & .01        \nl
 $(3.5) [1.5] $& 1 &   59718.13    & .01    \nl \nl
 $(3.5) [0.5] $& 1 &   59718.41$^P$& .05        \nl
 $(3.5) [0.5] $& 0 &   59718.41$^P$& .05    \nl \nl
 $(2.5) [6.5] $& 7 &   59999.211   & .007       \nl
 $(2.5) [6.5] $& 6 &   59999.206   & .004   \nl \nl
 $(2.5) [5.5] $& 6 &   60001.343   & .004       \nl
 $(2.5) [5.5] $& 5 &   60001.337   & .004   \nl \nl
 $(2.5) [4.5] $& 5 &   60001.575   & .004       \nl
 $(2.5) [4.5] $& 4 &   60001.56    & .01    \nl \nl
 $(2.5) [3.5] $& 4 &   60000.71$^S$& .01        \nl
 $(2.5) [3.5] $& 3 &   60000.71$^S$& .01    \nl \nl
 $(2.5) [2.5] $& 3 &   59999.394   & .005       \nl
 $(2.5) [2.5] $& 2 &   59999.390   & .006   \nl \nl
 $(2.5) [1.5] $& 2 &   59998.124   & .01        \nl
 $(2.5) [1.5] $& 1 &   59998.12    & .01    \nl \nl
 $(1.5) [5.5] $& 6 &   60193.670   & .007       \nl
 $(1.5) [5.5] $& 5 &   60193.666   & .005   \nl \nl
 $(1.5) [4.5] $& 5 &   60197.940   & .006       \nl
 $(1.5) [4.5] $& 4 &   60197.937   & .003   \nl \nl
 $(1.5) [3.5] $& 4 &   60196.429   & .007       \nl
 $(1.5) [3.5] $& 3 &   60196.42    & .01    \nl \nl
 $(1.5) [2.5] $& 3 &   60192.14$^S$& .01        \nl
 $(1.5) [2.5] $& 2 &   60192.14    & .01    \nl \nl
 $(0.5) [4.5] $& 5 &   60309.69    & .01        \nl
 $(0.5) [4.5] $& 5 &   60309.70    & .01    \nl \nl
 $(0.5) [3.5] $& 4 &   60309.711   & .006       \nl
 $(0.5) [3.5] $& 3 &   60309.717   & .005       \nl
\nl
\tablenotetext{a}{ Levels are marked in JK notation: $(J_c)[K].$}
\tablenotetext{P}{ Predicted level value (see text). }
\tablenotetext{S}{ Level value derived from solar spectrum}
\end{planotable}

\clearpage
\begin{planotable}{@{}r@{}l@{}c@{}lll@{}r@{}rrr@{}l@{}r@{--}rl}
\tablenum{2}
\tablewidth{0pt}
\tablecaption{$4f - 5g$ lines observed in Fe~I}
\tablehead{               \multicolumn{5}{c}{Laboratory Data}
                         &\multicolumn{3}{c}{Solar Data}
                         &\multicolumn{5}{c}{Identification}
                         &\multicolumn{1}{c}{Comments\tablenotemark{g}}\\
                          \multicolumn{5}{c}{\hrulefill}
                         &\multicolumn{3}{c}{\hrulefill}
                         &\multicolumn{5}{c}{\hrulefill}\\
                          \multicolumn{2}{c}{I\tablenotemark{a}}
                         &\multicolumn{1}{c}{width\tablenotemark{b}}
                         &\multicolumn{1}{c}{$\lambda_{air}$}
                         &\multicolumn{1}{c}{$\sigma_l$}
                         &\multicolumn{1}{c}{$\sigma_s$}
                         &\multicolumn{1}{c}{I}
                         &\multicolumn{1}{r}{log(gf)\tablenotemark{d}}
                         &\multicolumn{1}{c}{$\delta \sigma$\tablenotemark{c}}
                         &\multicolumn{2}{c}{log(gf)\tablenotemark{e}}
                         &\multicolumn{2}{c}{Transition\tablenotemark{f}}\\
                        &&\multicolumn{1}{c}{/mK}
                         &\multicolumn{1}{c}{/ \AA}
                         &\multicolumn{1}{c}{/cm$^{-1}$}
                         &\multicolumn{1}{c}{/cm$^{-1}$}
                         &\multicolumn{1}{c}{/mK}
                        &&\multicolumn{1}{c}{/mK}
                       &&&\multicolumn{1}{c}{$4f$}
                         &\multicolumn{1}{c}{$5g$}
}
\startdata
 1.0&   & 56& 39274.98& 2545.456&.462& 1.98 & 0.11& -3& 0.16 &   & $(4.5)
[1.5]_1^{\circ} $&$(4.5) [1.5]_1$& $*$     \nl
    &   &   & 39274.98& 2545.456&    &      &-0.56& -6&-0.51 &   & $(4.5)
[1.5]_1^{\circ} $&$(4.5) [1.5]_2$& $*$     \nl
 1.0&   & 31& 39256.54& 2546.652&.636& 1.56 &-0.37& 16&-0.24 &   & $(4.5)
[1.5]_1^{\circ} $&$(4.5) [0.5]_1$&         \nl
    &   &   & 39256.54& 2546.652&    &      &-0.10&  1& 0.03 &   & $(4.5)
[1.5]_1^{\circ} $&$(4.5) [0.5]_0$&         \nl
 2.0&   & 81& 39245.13& 2547.392&.392& 2.60 &-0.67&  9&-0.52 &   & $(4.5)
[1.5]_2^{\circ} $&$(4.5) [1.5]_1$& $*$       \nl
    &   &   & 39245.13& 2547.392&    &      & 0.27&  6& 0.43 &   & $(4.5)
[1.5]_2^{\circ} $&$(4.5) [1.5]_2$& $*$       \nl
 2.5&   & 64& 39227.14& 2548.561&.567& 2.38 & 0.27&  1& 0.42 &   & $(4.5)
[1.5]_2^{\circ} $&$(4.5) [0.5]_1$&         \nl
 4.8&   & 59& 39166.76& 2552.489&.486& 4.97 & 0.34&  3& 0.44 &   & $(4.5)
[7.5]_8^{\circ} $&$(4.5) [7.5]_8$& $*$       \nl
    &   &   & 39166.76& 2552.489&.496&      & 0.28& -7& 0.38 &   & $(4.5)
[7.5]_7^{\circ} $&$(4.5) [7.5]_7$& $*$       \nl
 1.5&   & 37& 39147.83& 2553.724&.724& 2.08 & 0.21&  1& 0.32 &   & $(4.5)
[2.5]_2^{\circ} $&$(4.5) [3.5]_3$& $*$     \nl
 5.4&   & 47& 39119.70& 2555.560&.563& 6.78 & 0.85&  1& 0.94 &   & $(1.5)
[3.5]_3^{\circ} $&$(1.5) [4.5]_4$& bl. OH \nl
 2.4&   & 47& 39116.28& 2555.784&.782& 2.98 & 0.37&  1& 0.47 &   & $(4.5)
[2.5]_2^{\circ} $&$(4.5) [2.5]_2$& $*$       \nl
 1.8&   & 37& 39113.03& 2555.996&.996& 2.95 & 0.37&  1& 0.46 &   & $(4.5)
[2.5]_3^{\circ} $&$(4.5) [3.5]_4$& $*$       \nl
 7.4&   & 49& 39101.86& 2556.726&.726& 8.56 & 0.97&  1& 1.04 &   & $(1.5)
[3.5]_4^{\circ} $&$(1.5) [4.5]_5$& $*$       \nl
38.2&   & 56& 39098.46& 2556.948&.947&18.5  & 0.99&  2& 1.26 &   & $(4.5)
[7.5]_7^{\circ} $&$(4.5) [8.5]_8$& $*$       \nl
    &   &   & 39098.46& 2556.948&    &      & 1.04&  1& 1.31 &   & $(4.5)
[7.5]_8^{\circ} $&$(4.5) [8.5]_9$& $*$       \nl
 1.1&   & 43& 39089.30& 2557.547&.55 & 2.0  & 0.17&  3& 0.17 &   & $(4.5)
[2.5]_2^{\circ} $&$(4.5) [1.5]_1$&  bl. Si \nl
    &   &   & 39089.30& 2557.547&    &      &-0.82&  0&-0.81 &   & $(4.5)
[2.5]_2^{\circ} $&$(4.5) [1.5]_2$&  bl. Si \nl
 3.4&   & 65& 39081.53& 2558.056&.05 & 5.16 & 0.63&  0& 0.63 &   & $(4.5)
[2.5]_3^{\circ} $&$(4.5) [2.5]_3$&         \nl
 1.0&   & 54& 39059.73& 2559.484&.48 & 0.78 &-0.12& -5&-0.17 &   & $(1.5)
[2.5]_2^{\circ} $&$(1.5) [2.5]_2$&         \nl
 1.2&   & 46& 39054.63& 2559.818&.82 & 3.18 & 0.40& -3& 0.35 &   & $(4.5)
[2.5]_3^{\circ} $&$(4.5) [1.5]_2$&         \nl
 3.0&$*$& 41& 39020.17& 2562.079&.095&12.22 & 0.77&  0& 0.87 &   & $(2.5)
[3.5]_3^{\circ} $&$(2.5) [4.5]_4$&         \nl
    &   &   & 39020.17& 2562.079&    &      &-0.44&  5&-0.34 &   & $(3.5)
[0.5]_0^{\circ} $&$(3.5) [1.5]_1$&         \nl
 8.0&$*$& 45& 39019.59& 2562.117&.127&      & 0.85& -1& 1.01 &   & $(2.5)
[4.5]_4^{\circ} $&$(2.5) [5.5]_5$&         \nl
 0.6&   & 34& 39018.24& 2562.206&.21 & 0.97 &-0.05&  6&\nodata&   & $(^6D)6p
^5P_2^{\circ} $&$(2.5) [2.5]_2$&         \nl
 2.7&   & 82& 39015.94& 2562.356&.349& 2.23 & 0.31& 15& 0.31 &   & $(2.5)
[4.5]_4^{\circ} $&$(2.5) [4.5]_4$& $*$     \nl
    &   &   & 39015.94& 2562.356&    &      &     &  0&\nodata&   & $(2.5)
[4.5]_4^{\circ} $&$(2.5) [4.5]_5$& $*$     \nl
10.9&   & 54& 38999.81& 2563.416&.420&10.5  & 1.05& -2& 1.10 &   & $(2.5)
[4.5]_5^{\circ} $&$(2.5) [5.5]_6$& $*$       \nl
 6.2&$*$& 43& 38998.05& 2563.532&.52 & 4.12 & 0.59&-11&\nodata&   & $(2.5)
[3.5]_4^{\circ} $&$(2.5) [5.5]_5$& $ \Delta K$=2 \nl
    &   &   & 38998.05& 2563.532&    &      & 0.59&  2&\nodata&   & $(^6D)6p
^5P_2^{\circ} $&$(2.5) [3.5]_3$&         \nl
 1.0&$*$& 46& 38996.23& 2563.651&.655& 1.86 & 0.20&  1& 0.37 &   & $(2.5)
[4.5]_5^{\circ} $&$(2.5) [4.5]_5$&         \nl
 8.1&$*$& 47& 38994.34& 2563.776&.779&11.4  & 0.87& -5& 0.97 &   & $(2.5)
[3.5]_4^{\circ} $&$(2.5) [4.5]_5$& $*$       \nl
    &   &   & 38994.34& 2563.776&    &      & 0.69&  4& 0.79 &   & $(1.5)
[2.5]_2^{\circ} $&$(1.5) [3.5]_3$& $*$       \nl
 1.1&$*$& 42& 38993.15& 2563.854&.85 & 0.94 &-0.10&-11& 0.21 &$*$& $(3.5)
[1.5]_2^{\circ} $&$(3.5) [1.5]_2$&         \nl
 1.6&$*$& 43& 38992.14& 2563.920&.90 & 2.09 & 0.25&  8& 0.27 &   & $(3.5)
[6.5]_6^{\circ} $&$(3.5) [6.5]_6$&         \nl
18.7&   & 45& 38979.21& 2564.771&.770&15.3  & 1.08& -2& 1.21 &   & $(3.5)
[6.5]_6^{\circ} $&$(3.5) [7.5]_7$& $*$       \nl
    &   &   & 38979.21& 2564.771&    &      & 0.85&  1& 0.97 &   & $(0.5)
[3.5]_3^{\circ} $&$(0.5) [4.5]_4$& $*$       \nl
 2.7&   & 47& 38972.11& 2565.238&.238& 5.06 & 0.61&  1& 0.61 &   & $(4.5)
[3.5]_3^{\circ} $&$(4.5) [4.5]_4$&   bl. Si\nl
 7.5&   & 43& 38970.59& 2565.338&.338& 8.77 & 0.26& -9& 0.30 &$*$& $(3.5)
[2.5]_2^{\circ} $&$(3.5) [2.5]_2$& $*$       \nl
    &   &   & 38970.59& 2565.338&    &      & 0.88&  1& 0.92 &   & $(1.5)
[2.5]_3^{\circ} $&$(1.5) [3.5]_4$& $*$       \nl
 1.0&   & 42& 38968.21& 2565.495&.500& 1.33 & 0.05&  0& 0.25 &   & $(3.5)
[1.5]_1^{\circ} $&$(3.5) [2.5]_2$&   bl. OH\nl
13.3&   & 44& 38966.47& 2565.609&.610&12.2  & 1.10&  0& 1.27 &   & $(3.5)
[6.5]_7^{\circ} $&$(3.5) [7.5]_8$& $*$       \nl
 3.2&   & 53& 38964.72& 2565.724&.723& 2.45 & 0.08&  6& 0.07 &$*$& $(3.5)
[2.5]_2^{\circ} $&$(3.5) [1.5]_2$&         \nl
    &   &   & 38964.72& 2565.724&    &      &-0.05&  6&-0.06 &$*$& $(3.5)
[2.5]_2^{\circ} $&$(3.5) [1.5]_1$&         \nl
 2.1&   & 46& 38955.22& 2566.351&.347& 5.33 & 0.74&  1& 0.84 &   & $(0.5)
[2.5]_2^{\circ} $&$(0.5) [3.5]_3$& $*$       \nl
 3.4&   & 38& 38952.60& 2566.523&.530&13.3  & 0.57&  1& 0.72 &   & $(4.5)
[3.5]_4^{\circ} $&$(4.5) [4.5]_5$&         \nl
 8.1&   & 48& 38952.03& 2566.561&.550&      & 0.20& 12& 0.37 &   & $(2.5)
[2.5]_3^{\circ} $&$(2.5) [2.5]_3$&         \nl
    &   &   & 38952.03& 2566.561&.570&      & 0.92&  0& 1.07 &   & $(0.5)
[3.5]_4^{\circ} $&$(0.5) [4.5]_5$&         \nl
 2.6&   & 51& 38951.54& 2566.593&.582&      &-0.07& 10&-0.22 &   & $(0.5)
[3.5]_4^{\circ} $&$(0.5) [3.5]_4$&         \nl
 6.4&   & 49& 38950.49& 2566.662&.66 & 6.68 & 0.85&  1& 0.94 &   & $(0.5)
[2.5]_3^{\circ} $&$(0.5) [3.5]_4$&         \nl
 3.0&   & 47& 38942.22& 2567.207&.209& 3.90 & 0.11&  0& 0.61 &   & $(4.5)
[3.5]_3^{\circ} $&$(4.5) [3.5]_3$&         \nl
 2.1&   & 47& 38941.63& 2567.246&    &      & 0.30&  0& 0.76 &   & $(3.5)
[3.5]_3^{\circ} $&$(3.5) [4.5]_4$&         \nl
 2.6&   & 50& 38937.99& 2567.485&.483& 2.93 & 0.40&  2& 0.70 &   & $(3.5)
[2.5]_3^{\circ} $&$(3.5) [3.5]_4$&         \nl
 4.7&   & 47& 38935.97& 2567.619&.62 & 3.57 & 0.49&  0& 0.51 &   & $(3.5)
[3.5]_3^{\circ} $&$(3.5) [3.5]_3$& $*$     \nl
 3.1&   & 42& 38935.01& 2567.683&.680& 4.33 & 0.58&  0&\nodata&   & $(3.5)
[3.5]_4^{\circ} $&$(3.5) [5.5]_5$& $*, \Delta K$=2 \nl
15.4&$*$& 37& 38932.78& 2567.829&.830& 9.23 & 0.66&  2& 0.88 &$*$& $(3.5)
[3.5]_4^{\circ} $&$(3.5) [4.5]_5$& bl. in lab\nl
 1.1&$*$& 23& 38932.16& 2567.870&.860&      & 0.65&  5& 0.83 &   & $(2.5)
[2.5]_3^{\circ} $&$(2.5) [3.5]_4$&         \nl
 3.8&$*$& 66& 38931.37& 2567.922&.92 & 2.50 & 0.33&  0& 0.56 &   & $(3.5)
[2.5]_3^{\circ} $&$(3.5) [2.5]_3$&         \nl
 3.8&$*$& 98& 38929.41& 2568.052&.060& 3.25 & 0.45& -9&-0.23 &   & $(3.5)
[3.5]_3^{\circ} $&$(3.5) [2.5]_2$& $*$     \nl
    &   &   & 38929.41& 2568.052&    &      & 0.  &-14&\nodata&   & $(3.5)
[3.5]_3^{\circ} $&$(3.5) [2.5]_3$& $*$     \nl
 0.9&   & 24& 38926.95& 2568.214&.200& 1.83 & 0.19& -2& 0.61 &$*$& $(3.5)
[3.5]_4^{\circ} $&$(3.5) [3.5]_4$&         \nl
 1.7&   & 38& 38926.04& 2568.274&.280& 2.10 & 0.22&  0& 0.20 &$*$& $(2.5)
[2.5]_2^{\circ} $&$(2.5) [2.5]_2$&         \nl
    &   &   & 38926.04& 2568.274&    &      &-0.58& -4&-0.75 &$*$& $(2.5)
[2.5]_2^{\circ} $&$(2.5) [2.5]_3$&         \nl
 2.3&   & 23& 38922.75& 2568.491&.492& 5.05 & 0.52&  0& 0.72 &   & $(4.5)
[3.5]_4^{\circ} $&$(4.5) [3.5]_4$& $*$     \nl
    &   &   & 38922.75& 2568.491&    &      &-0.12& -2& 0.06 &   & $(4.5)
[3.5]_4^{\circ} $&$(4.5) [3.5]_3$& $*$     \nl
 2.7&   & 47& 38914.93& 2569.007&.011& 2.93 & 0.40& -4& 0.45 &   & $(3.5)
[5.5]_5^{\circ} $&$(3.5) [5.5]_5$& $*$       \nl
 9.7&   & 48& 38910.56& 2569.296&.299&10.22 & 1.00& -2& 1.08 &   & $(3.5)
[5.5]_5^{\circ} $&$(3.5) [6.5]_6$& $*$       \nl
 2.5&   & 43& 38905.52& 2569.629&.625& 2.63 & 0.35&  5& 0.93 &$*$& $(3.5)
[4.5]_4^{\circ} $&$(3.5) [5.5]_5$& $*$       \nl
 8.3&   & 39& 38903.47& 2569.764&.774& 4.73 & 0.62&-12& 0.54 &$*$& $(3.5)
[4.5]_4^{\circ} $&$(3.5) [4.5]_4$& bl. in lab\nl
 2.1&   & 43& 38899.57& 2570.021&.024& 3.28 & 0.45& -3& 0.55 &   & $(3.5)
[5.5]_6^{\circ} $&$(3.5) [5.5]_6$&         \nl
11.8&   & 55& 38898.19& 2570.113&.112&10.50 & 1.04&  1& 1.14 &   & $(2.5)
[5.5]_5^{\circ} $&$(2.5) [6.5]_6$& $*$       \nl
13.1&   & 57& 38895.16& 2570.313&.313&10.15 & 1.00& -2& 1.15 &   & $(3.5)
[5.5]_6^{\circ} $&$(3.5) [6.5]_7$& $*$       \nl
 2.7&   & 41& 38891.58& 2570.550&.553& 2.71 & 0.24& -2& 0.27 &   & $(4.5)
[3.5]_4^{\circ} $&$(4.5) [2.5]_3$& $*$       \nl
    &   &   & 38891.58& 2570.550&    &      & 0.14& -1& 0.20 &   & $(2.5)
[1.5]_2^{\circ} $&$(2.5) [1.5]_2$& $*$       \nl
 8.9&   & 52& 38884.60& 2571.011&.011&21.0  & 1.42&  0& 1.02 &   & $(3.5)
[4.5]_5^{\circ} $&$(3.5) [5.5]_6$&  bl. Si \nl
16.9&   & 54& 38882.47& 2571.152&.153&13.5  & 0.44& -2& 0.59 &   & $(3.5)
[4.5]_5^{\circ} $&$(3.5) [4.5]_5$& $*$     \nl
    &   &   & 38882.47& 2571.152&    &      & 1.06&  0& 1.21 &   & $(2.5)
[5.5]_6^{\circ} $&$(2.5) [6.5]_7$& $*$     \nl
 6.0&   & 34& 38873.56& 2571.741&.741& 8.78 & 0.97&  0& 1.07 &   & $(1.5)
[4.5]_4^{\circ} $&$(1.5) [5.5]_5$& $*$       \nl
 4.0&   & 53& 38872.34& 2571.822&.82 & 4.01 & 0.57&  1& 0.68 &   & $(2.5)
[1.5]_2^{\circ} $&$(2.5) [2.5]_3$&         \nl
 0.6&   & 19& 38865.90& 2572.248&.24 & 2.57 & 0.37&  5& 0.13 &   & $(2.5)
[5.5]_5^{\circ} $&$(2.5) [5.5]_5$&  bl. Si \nl
14.1&   & 33& 38856.33& 2572.882&.884& 8.77 & 0.92&  1& 1.15 &   & $(1.5)
[4.5]_5^{\circ} $&$(1.5) [5.5]_6$&  bl. in lab\nl
 4.0&   & 52& 38852.39& 2573.142&.142& 3.39 & 0.42&  0& 0.57 &   & $(4.5)
[6.5]_6^{\circ} $&$(4.5) [6.5]_6$&         \nl
 1.5&   & 50& 38850.23& 2573.286&.285& 1.75 & 0.20&  2& 0.20 &   & $(2.5)
[5.5]_6^{\circ} $&$(2.5) [5.5]_6$&         \nl
12.4&   & 53& 38827.83& 2574.770&.770&10.72 & 0.99&  0& 1.14 &   & $(4.5)
[6.5]_6^{\circ} $&$(4.5) [7.5]_7$& $*$       \nl
 5.8&   & 44& 38824.95& 2574.961&.962& 6.43 & 0.72& -1& 0.82 &$*$& $(4.5)
[4.5]_4^{\circ} $&$(4.5) [5.5]_5$& $*$       \nl
 8.2&   & 56& 38821.81& 2575.169&.170& 6.28 & 0.71& -1& 0.91 &   & $(4.5)
[4.5]_5^{\circ} $&$(4.5) [5.5]_6$& $*$       \nl
 2.8&   & 35& 38816.49& 2575.522&.519& 4.11 & 0.51&  3& 0.66 &   & $(4.5)
[6.5]_7^{\circ} $&$(4.5) [6.5]_7$& $*$       \nl
 4.2&   & 45& 38803.83& 2576.363&.359& 4.80 & 0.58&  5& 0.67 &$*$& $(4.5)
[4.5]_4^{\circ} $&$(4.5) [4.5]_4$& $*$     \nl
 6.2&   & 64& 38800.84& 2576.561&.564& 5.85 & 0.68& -3& 0.75 &   & $(4.5)
[4.5]_5^{\circ} $&$(4.5) [4.5]_5$&         \nl
 2.2&   & 46& 38799.23& 2576.668&.663& 3.02 &-0.17& 11& 0.00 &   & $(2.5)
[0.5]_1^{\circ} $&$(2.5) [1.5]_1$&         \nl
    &   &   & 38799.23& 2576.668&.   &      & 0.32&  7& 0.49 &   & $(2.5)
[0.5]_1^{\circ} $&$(2.5) [1.5]_2$&         \nl
15.6&   & 53& 38792.07& 2577.143&.143&12.18 & 1.06&  0& 1.20 &   & $(4.5)
[6.5]_7^{\circ} $&$(4.5) [7.5]_8$& $*$       \nl
10.1&   & 53& 38770.95& 2578.547&.547& 8.13 & 0.84&  0& 0.99 &   & $(4.5)
[5.5]_5^{\circ} $&$(4.5) [6.5]_6$& $*$       \nl
 4.0&   & 45& 38767.21& 2578.796&.796& 5.08 & 0.61&  0& 0.68 &   & $(4.5)
[5.5]_5^{\circ} $&$(4.5) [5.5]_5$& $*$       \nl
10.9&   & 49& 38737.93& 2580.745&.747& 9.86 & 0.94& -1& 1.06 &   & $(4.5)
[5.5]_6^{\circ} $&$(4.5) [6.5]_7$& $*$       \nl
 5.0&   & 48& 38734.17& 2580.996&.994& 5.94 & 0.68&  2& 0.72 &   & $(4.5)
[5.5]_6^{\circ} $&$(4.5) [5.5]_6$& $*$      \nl
 0.6&   & 17& 38713.34& 2582.385&.388& 1.28 &-0.02& -3&-0.02 &   & $(4.5)
[5.5]_6^{\circ} $&$(4.5) [4.5]_5$&  bl. Si \nl
\tablenotetext{a} {Intensity in arbitrary units. An asterisk marks unresolved
lines.
          In the case of unresolved multiple components (in laboratory and
solar spectra), the total
          intensity of the blend is only given in the row relative to the first
component.}
\tablenotetext{b} {Full width at half maximum of line in FT spectra. Units are
mK (1mK = 0.001 cm$^{-1}$)}
\tablenotetext{c} {Difference between laboratory wavenumber and that derived
from the energy levels in table 1.}
\tablenotetext{d} {log(gf) value derived from the solar line intensity (with
the solar equivalent width I in mK).}
\tablenotetext{e} {Calculated log(gf) value. An asterisk marks mixed levels for
which the calculation is unreliable.}
\tablenotetext{f} {The $3d^64s(^6D)4f$ configuration has been abbreviated as
$4f$, and $3d^64s(^6D)5g$ as $5g$.
          Levels are given in JK notation: (J$_c$)[K], where J$_c$ is the J
value of the parent level.}
\tablenotetext{g} {An asterisk marks a good solar line profile. Blends with
other solar lines are indicated.}
\end{planotable}

\clearpage

\begin{planotable}{rl@{}c@{}lrrr@{}l@{}r@{--}rl}
\tablenum{3}
\tablewidth{0pt}
\tablecaption{Additional $4f-5g$ lines derived from the energy levels and
            observed in the solar spectrum}
\tablehead{              &
                         &\multicolumn{3}{c}{Solar Data}
                         &\multicolumn{5}{c}{Identification}
                         &\multicolumn{1}{c}{Comments\tablenotemark{f}}\\
                          \multicolumn{2}{c}{\hrulefill}
                         &\multicolumn{3}{c}{\hrulefill}
                         &\multicolumn{5}{c}{\hrulefill}\\
                          \multicolumn{1}{c}{$\lambda_{air}$}
                         &\multicolumn{1}{c}{$\sigma$\tablenotemark{a}}
                         &\multicolumn{1}{c}{$\sigma_s$}
                         &\multicolumn{1}{c}{I}
                         &\multicolumn{1}{r}{log(gf)$_s$\tablenotemark{b}}
                         &\multicolumn{1}{c}{$\delta \sigma$\tablenotemark{c}}
                         &\multicolumn{2}{c}{log(gf)$_c$\tablenotemark{d}}
                         &\multicolumn{2}{c}{Transition\tablenotemark{e}}\\
                          \multicolumn{1}{c}{/ \AA}
                         &\multicolumn{1}{c}{/cm$^{-1}$}
                         &\multicolumn{1}{c}{/cm$^{-1}$}
                         &\multicolumn{1}{c}{/mK}
                        &&\multicolumn{1}{c}{/mK}
                       &&&\multicolumn{1}{c}{$4f$}
                         &\multicolumn{1}{c}{$5g$}
                         &
  }
\startdata
 39302.12& 2543.698&.690& 0.86 &-0.17&  8&-0.13 &   & $(4.5) [1.5]_1^{\circ}
$&$(4.5) [2.5]_2$&         \nl
 39142.79& 2554.045&.047& 1.15 & 0.05& -2&-0.01 &   & $(1.5) [3.5]_3^{\circ}
$&$(1.5) [3.5]_3$& $*$     \nl
 39124.99& 2555.214&.224& 1.51 & 0.17&-10& 0.11 &   & $(1.5) [3.5]_4^{\circ}
$&$(1.5) [3.5]_4$& Ne      \nl
 39035.87& 2561.043&.053& 0.98 &-0.02& -5& 0.00 &   & $(1.5) [2.5]_3^{\circ}
$&$(1.5) [2.5]_3$&         \nl
 39032.65& 2561.229&.230& 2.02 & 0.27& -1& 0.30 &   & $(2.5) [3.5]_3^{\circ}
$&$(2.5) [3.5]_3$& $*$     \nl
 39007.02& 2562.916&.908& 2.42 & 0.35&  8& 0.43 &   & $(2.5) [3.5]_4^{\circ}
$&$(2.5) [3.5]_4$& $*$ Ne  \nl
 38945.30& 2567.004&.010& 1.06 &-0.01& -6&-0.99 &$*$& $(2.5) [2.5]_2^{\circ}
$&$(2.5) [1.5]_1$& $*$     \nl
 38945.24& 2567.008&    &      &     & -2&-1.61 &$*$& $(2.5) [2.5]_2^{\circ}
$&$(2.5) [1.5]_2$& $*$     \nl
 38812.82& 2575.766&.774& 0.68 &-0.29& -8&-0.29 &   & $(4.5) [6.5]_7^{\circ}
$&$(4.5) [5.5]_6$& $*$     \nl
 38809.11& 2576.017&.015& 0.85 &-0.09&  2&-0.09 &   & $(1.5) [4.5]_4^{\circ}
$&$(1.5) [4.5]_4$&         \nl
 38874.28& 2578.326&.326& 1.53 & 0.06&  2& 0.06 &   & $(4.5) [4.5]_4^{\circ}
$&$(4.5) [3.5]_3$& $*$     \nl
\tablenotetext{a}{ Wavenumber and wavelengths derived from the energy levels in
          table 1.}
\tablenotetext{b}{ log(gf) value derived from solar spectrum. }
\tablenotetext{c}{ Difference between wavenumber in solar spectrum and that in
          column 2. }
\tablenotetext{d}{ Calculated log(gf) value. An asterisk marks mixed levels for
          which the calculation is unreliable. }
\tablenotetext{e}{ The  $3d^64s(^6D)4f$  configuration has been abbreviated as
          $4f$, and  $3d^64s(^6D)5g$  as $5g$. Levels are given in JK
          notation: (J$_c$)[K], where J$_c$ is the J value of the parent
          level. }
\tablenotetext{f}{ An asterisk marks a good solar line profile. `Ne' indicates
          the line is masked by a neon line in the laboratory spectrum. }
\end{planotable}

\twocolumn

\pagebreak

\section*{Figures}

\begin{enumerate}

\item Region of $4f-5g$ transitions in Fe~I. The upper section gives the
      solar spectrum (in absorption), and the lower section the
      laboratory spectrum (in emission). The two broad lines at
      2555.13~\cm\ and 2562.78~\cm\ in the laboratory spectrum are neon
      lines. Some of the stronger $4f-5g$ lines are marked by dotted lines.

\item Partial energy level diagram of Fe~I showing the \5g energy levels
      arranged in the JK coupling representation. The ordinate is
      $h=\frac{1}{2}\{K(K+1) - J_c(J_c+1)- l(l+1)\}$. The curves are
      parabolas fitted to the observed energy levels (circles) for each
      value of $J_c$.

\item Comparison of observed solar equivalent widths and laboratory
      intensities for \4f -- \5g lines. The point marked `bl. Si' is a
      blend of a $4f-5g$ line and a Si line in the solar spectrum.

\item Comparison of observed and calculated solar line spectra with
      enhancement factors of 2 (top), 2.5 (middle), and 3 (bottom) for
      the collisional damping constant. The effect of an increase of the
      enhancement factor is clearly seen in Fe~I \4f-5g\ line profiles.

\item Section of the solar spectrum(top) with a fitted synthetic
      spectrum indicated by dashed lines. The laboratory spectrum is
      shown below. The majority of lines in this section are due to Fe I
      and, with a few exceptions, can be assigned to the \4f$-$\5g
      supermultiplet.

\item Comparison of calculated gf-values and gf-values derived from the
      solar spectrum for \4f -- \5g lines. Only the best, unblended
      solar lines (marked $*$ in the last column of table 2) are
      included.

\end{enumerate}

\end{document}